# Enhanced Magnetism & Time - Stable Remanence at the Interface of Hematite and Carbon Nanotubes

Aakanksha Kapoor[1], Arka Bikash Dey[2], Charu Garg[1] and Ashna Bajpai[1-3]

1. Department of Physics, Indian Institute of Science Education and Research, Pune-411008, India.

2. Saha Institute of Nuclear Physics, 1/AF Bidhannagar, Kolkata, India.

3. Centre for Energy Science, Indian Institute of Science Education and Research, Dr. Homi Bhabha Road, Pune, Maharashtra-411008, India.

Email: ashna@iiserpune.ac.in

The interface of two dissimilar materials is well known for surprises in condensed matter, and provides avenues for rich physics as well as seeds for future technological advancements. We present some exciting magnetization (M) and remnant magnetization ($\mu$) results, which conclusively arise at the interface of two highly functional materials, namely the graphitic shells of a carbon nanotube (CNT) and $\alpha$-$Fe_2O_3$, a Dzyaloshinskii-Moriya Interaction (DMI) driven weak ferromagnet (WFM) and piezomagnet (PzM). We show that the encapsulation inside CNT leads to a very significant enhancement in M and correspondingly in $\mu$, a *time- stable* part of the remanence, exclusive to the WFM phase. Up to 70% of in-field magnetization is retained in the form of $\mu$ at the room temperature. Lattice parameter of CNT around the Morin transition of the encapsulate exhibits a clear anomaly, confirming the novel interface effects. Control experiments on bare $\alpha$-$Fe_2O_3$ nanowires bring into fore that the weak ferromagnets such as $\alpha$-$Fe_2O_3$ as are not as weak, as far as their remanence and its stability with time is concerned, and encapsulation inside CNT leads to a substantial enhancement in these functionalities.

Hematite (or $\alpha$-Fe$_2$O$_3$) is an earth abundant and environment-friendly oxide, generally considered as a menace, for its appearance as common rust over elemental Fe, but technologically, it is well known for a very diverse range of applications.[1] However, the observation of WFM[2, 3] nearly six decades ago in hematite and its connection to spin orbit coupling (SOC) has had profound implications in the field of spintronics. A variety of non-trivial topological spin structures in chiral magnets stabilizing through DMI/SOC have triggered new research areas such as antiferromagnetic spintronics and spin orbitronics[4,9].

Generation of WFM in $\alpha$-Fe$_2$O$_3$ is due to a slight canting of its inherent AFM sublattice [2, 3] which persists from 950 K (T$_N$) down to 265 K, well-known Morin Transition temperature (T$_M$). Below T$_M$, the spins turn from "*a*" axis to "*c*" axis (rhombohedral unit cell in hex setting) and the canting vanishes for hematite. Many of such canted AFM, either DMI driven or systems in which canting takes place due to other mechanism, are also known to exhibit the phenomenon of piezomagnetism.[10,14] Here, PzM implies that magnetization can be tuned by stress alone. Another important point is the occurrence of WFM which is concurrent with PzM as theoretically predicted [10] and experimentally observed[11,14]. Dzyaloshinskii also showed that the spin canting effect is larger for compounds with smaller T$_N$[2]. Thus, WFM / PzM is seen to be the weakest in $\alpha$-Fe$_2$O$_3$ with T$_N$ ~950 K as compared to MnCO$_3$ or NiCO$_3$ (T$_N$ below 50 K).

This work centers around remanence μ, which, in general, is an important parameter for any magnetic material for a variety of practical applications related to permanent magnets, soft or hard, relate to this quantity[15,17]. In addition, it is an important tool for probing fundamental magnetic interactions in conventional long range order (LRO) as well as complex magnetic systems, including frustrated, nano and core shell magnets.[17,20] These systems are better identified by simple remanence measurements, which carry crucial information about the underlying magnetic interaction associated with each of these distinct phases. Recently, we have explored in a number of DMI driven WFM & PzM using SQUID magnetometry.[21] The key result there is the observation of highly *time-stable* μ along with some other features that exclusively connect it to WFM phase.[21]

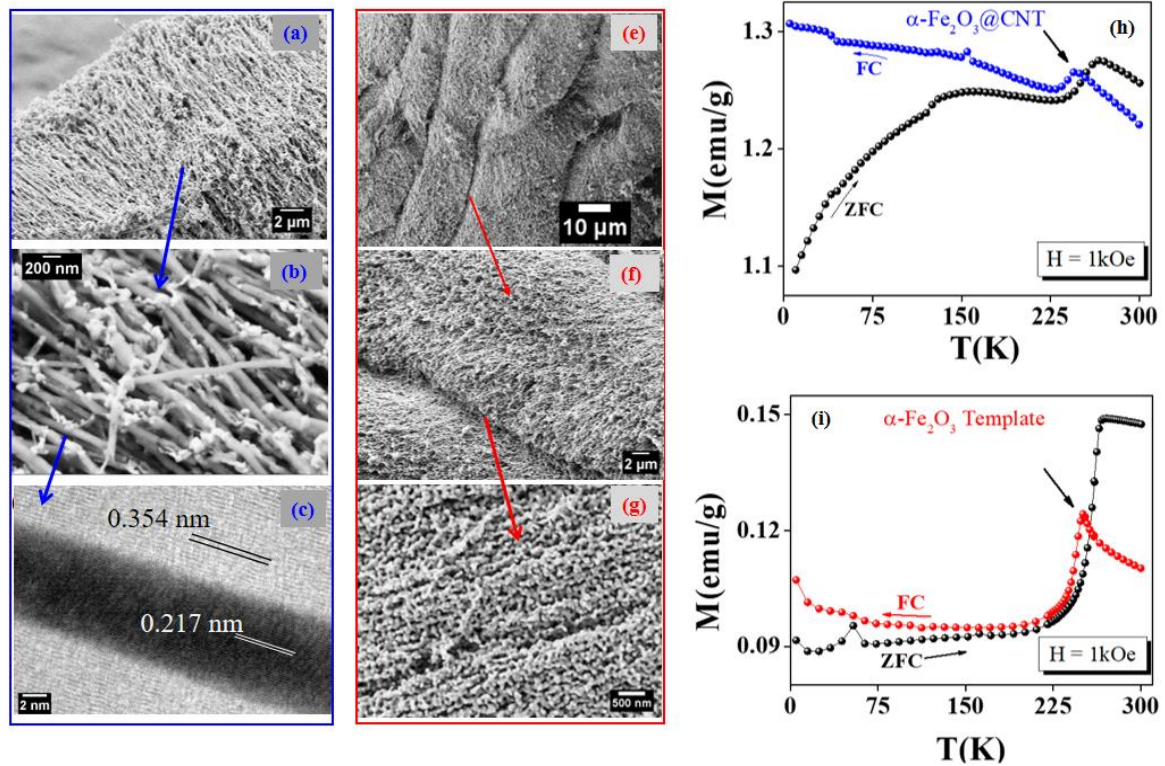

**Figure 1**: *(a)-(b) are SEM images of α-Fe$_2$O$_3$@CNT at different magnifications. (c) TEM image of α-Fe$_2$O$_3$@CNT depicting well-formed graphitic shells of CNT and the oxide encapsulate. (e) &(g) are the SEM images for the oxide template . (h) shows M vs T in a typical FC (blue dots) and ZFC (black dots) for the sample α-Fe$_2$O$_3$@CNT measured 1kOe (i) shows the same for the oxide template.*

On a general note, magnetic oxides such as hematite are well known to exhibit a remarkably wide range of functional properties, [22,23] device fabrication still remains a challenge due to a number of practical issues. In this light, the oxide encapsulation inside carbon nanotubes is potentially beneficial for a number of reasons. Owing to a wide diversity in the electronic and magnetic ground states of the encapsulate [22,30] together with unprecedented mechanical, electrical and thermal properties of the CNT,[24] these hybrids are potential candidates for novel interface effects.[25-28] In this work, we present experimental results that indicate that encapsulation of α-Fe$_2$O$_3$ inside CNT is the most efficient way to enhance the magnitude of the DMI driven WFM and consequently PzM.

The CNT with α-Fe$_2$O$_3$ encapsulate has been obtained in aligned forest form[31]. This sample is referred to as α-Fe$_2$O$_3$@CNT (**Figure 1a-1b & Text S1**). Scanning Electron Microscope (SEM) images are acquired using ZEISS ULTRA plus field-emission SEM Transmission Electron Microscope (TEM) image shown in **Figure 1 (c)** is obtained using FEI-TECNAI microscope. α-Fe$_2$O$_3$@CNT sample is further annealed at higher temperatures so as to intentionally remove the CNT and form the oxide template in the same morphology, (**Figure 1e-1g**) This sample is referred to as α-Fe$_2$O$_3$ template (**Text S1**). The temperature variation of synchrotron XRD from 20- 300 K has been conducted in BL-18 beamline, Photon factory, Japan and fitted using Rietveld Profile Refinement[32] (**Figure S1**).

The bulk magnetization measurements have been performed using Superconducting Quantum Interface Device (SQUID) magnetometer from Quantum Design. The M vs T measurements in presence of a fix magnetic field (H) have been taken in routine Zero Field Cooled (ZFC) and Field Cooled (FC) cycles, The corresponding M has been referred as M$_{ZFC}$ and M$_{FC}$ respectively. The remnant state is prepared following either the FC (or ZFC) protocol. This basically involves cooling (or heating) the sample in presence of H from above (or from below) the T$_M$ of α-Fe$_2$O$_3$ and eventually switching off H at 5K (or at 300K) for measuring μ$_{FC}$ (or μ$_{ZFC}$). It is to be emphasized that the μ is strictly measured in H = 0 condition. The H indicated in all the figures pertaining to μ is to convey its magnitude while preparing a remnant state.

**Enhanced Magnetization:** Magnetization as a function of temperature for α-Fe$_2$O$_3$@CNT in a typical FC (blue dots) and ZFC (black dots) cycles, measured at H~1kOe is shown in (**Figure 1h**). Same is shown for the α-Fe$_2$O$_3$ template (**Figure 1i**). Both the samples exhibit well pronounced $T_M$, intrinsic to the bulk α-Fe$_2$O$_3$. The shift in $T_M$ towards lower temperatures (as compared to bulk α-Fe$_2$O$_3$) as we the functional form of M vs T is consistent with the fact that α-Fe$_2$O$_3$ in both cases is in the form of nano wires. Thus, the broad features of magnetization and history effects (i.e. bifurcation in FC/ZFC cycles) can be attributed to nano scaling, consistent with previous reports.[33, 34] In conventional LRO or in complex magnetic systems $M_{FC} > M_{ZFC}$ is a frequently observed phenomenon.[17, 35] However, we observe the opposite, especially in the temperature region above $T_M$ (**Figure 1h&1i**). This is not a common occurrence, but it has been reported earlier.[36, 37] We shall come back to this issue in the latter part of the text.

The more prominent observation is about an order of magnitude enhancement in M between α-Fe$_2$O$_3$@CNT and the α-Fe$_2$O$_3$ template at each H. For instance, $M_{FC}$ = 1.3 emu/g (**Figure 1i**), whereas it is ~ 0.1 emu/g for the template (**Figure 1j**) for H =1 kOe and T= 5K. Assuming that the graphitic shells of CNT do not contribute magnetically in a conventional sense, the data presented in **Figure 1h&1i** are intriguing. The enhancement is seen in all H for α-Fe$_2$O$_3$@CNT as compared to the template (**Figure S2**). It is to be noted that for the template, the magnitude of M is similar to what one usually observes in α-Fe$_2$O$_3$ nano particles.[21, 33, 34] If this enhancement in M was arising only due to the nano scaling of α-Fe$_2$O$_3$ or if it was morphology related (aligned forest), the effect should have persisted in the template. Though, both the size effects & morphology certainly have a role to play as far as WFM is concerned, [21, 28], data in **Figure 1h&1i** suggest that the significant enhancement in M is related to the *interface* effects.

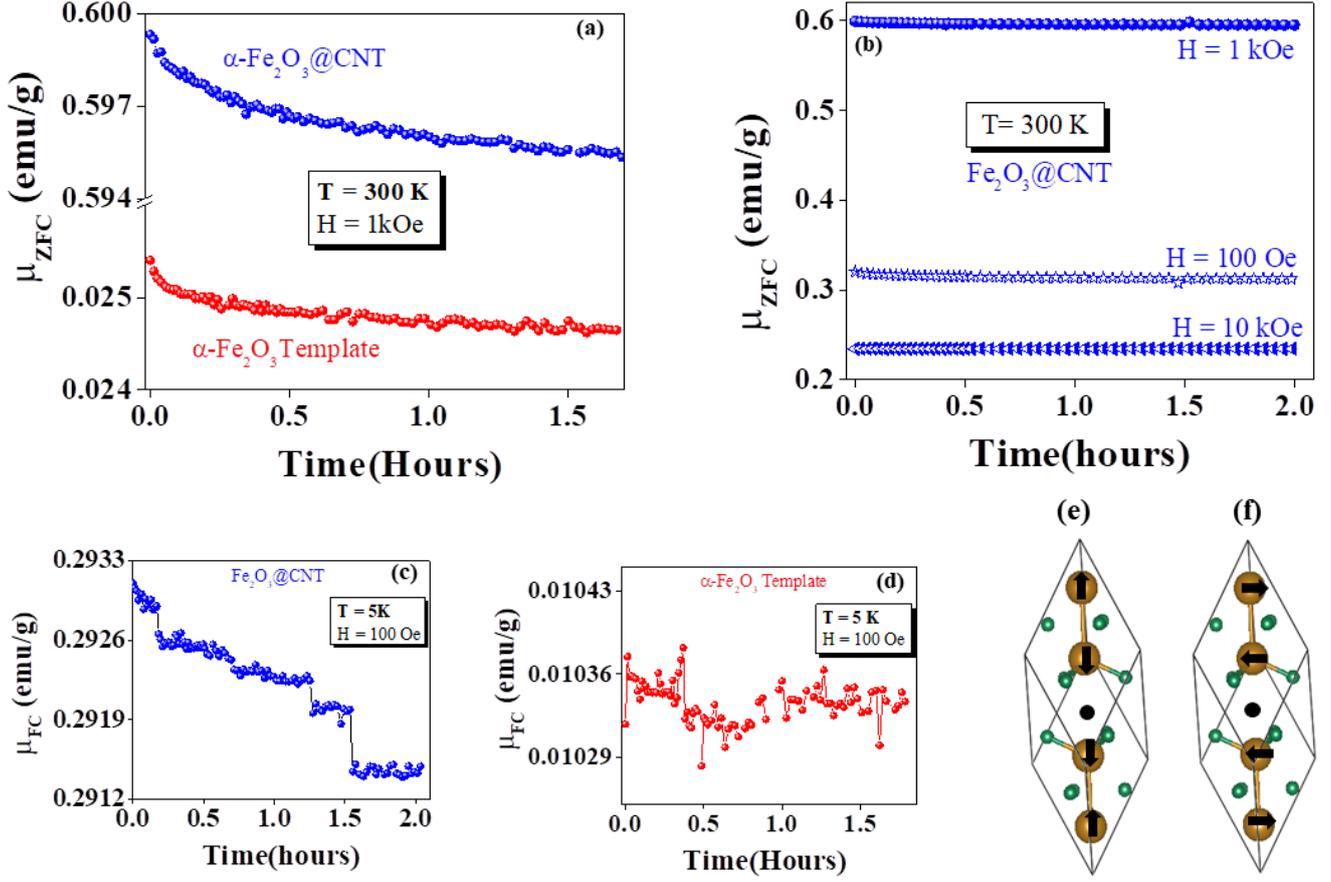

**Figure 2:** *(a)* $\mu_{ZFC}$ *vs time measured at 300K for α-Fe$_2$O$_3$@CNT (blue dots) and the oxide template (red dots) (b) shows $\mu_{ZFC}$ vs time for α-Fe$_2$O$_3$CNT at three different H, depicting its time-stable nature and its counter-intuitive H dependence. (c) & (d) show $\mu_{FC}$ vs time measured at 5 K for α-Fe$_2$O$_3$@CNT and the template respectively. Schematic of spin configuration in (e) pure AFM and (f) WFM state for rhombohedral unit cell of Hematite. The yellow and green spheres are the Fe and O ions respectively. The black dot is the inversion center.*

**Enhanced Remanence :** A more striking (and useful) result is the observation of a *time-stable* µ for both the samples, the magnitude of which is substantially enhanced in case of α-Fe$_2$O$_3$@CNT(**Figure 2a**). Here again, the magnitude of the $\mu_{ZFC}$ is at least an order of magnitude larger for α-Fe$_2$O$_3$@CNT (blue dots) than what is observed in the template (red dots). **Figure 2b** compares µ prepared at three different H for α-Fe$_2$O$_3$@CNT to depict its stability with time and its counter-intuitive H dependence. Similar features are observed in remnant state prepared following a FC protocol. The magnitude of $\mu_{FC}$ (~ 0.3 emu/g) is substantially larger (**Figure 2c**) as compared to the template (~ 0.01 emu/g) (**Figure 2d**), at 5K. It is also intriguing to note that the remanence is larger and more stable at 300 K than at 5K, even though thermal fluctuations should disrupt the remanence more at higher temperatures.

**Weak Ferromagnetism & Unusual Magnetization Dynamics :** Recently, investigating µ in a number of such canted AFM has enabled an understanding that these systems leave some unique footmarks in remanence, which are not evident in routine M vs T or M vs H cycles.[21] With special focus on α-Fe$_2$O$_3$, which shows both AFM and WFM phase across $T_M$ , the spin configuration is schematically shown for pure AFM (**Figure 2e**) and the WFM phase (**Figure 2f**), for which the spontaneous canting of the type $\vec{D}.\vec{S_i} \times \vec{S_j}$ is symmetry allowed, leading to a net ferromagnetic moment, in an otherwise AFM. It is to be recalled that the net *in-field magnetization* in an AFM depends on the energies including Zeeman, exchange and magnetocrystalline anisotropy.[15] In the case of WFM, the direction of net ferromagnetic moment due to the canted spins is an additional factor. In the light of above scenario, we discuss all the unusual results, the nature of FC/ZFC bifurcation in M vs T, the presence of a *time-stable* µ with a counter intuitive H dependence and enhanced magnitude of M & µ in case α-Fe$_2$O$_3$@CNT.

We first discuss the unusual history effects, $M_{FC} < M_{ZFC}$ (**Figure 1i-1j**). This feature is extremely sensitive to the exact sequence in which FC/ZFC runs are recorded for a particular H (**Text S2**). Spin canting in Hematite starts from $T_M$ and persisting right upto 960K, wherein a WFM to paramagnetic transition occurs. Considering spin configuration of WFM state from **Figure 2f**, it is clear that the direction of net FM moment associated with a canted AFM domain can be best modulated (w.r.t the direction of H) if H is applied from below $T_M$, just before the spin reorientation transition takes place for α-Fe$_2$O$_3$ (i.e. at the onset of spontaneous canting), leading to the larger $M_{ZFC}$. On the other hand, when H is applied from above the $T_M$, where spins are already spontaneously canted (with a net ferromagnetic moment), their contribution to the total magnetization is not entirely dictated by H. This is due to the fact that it may not be energetically favorable for small H to flip the direction of a WFM domain with net ferromagnetic moment pointing in unfavorable direction with respect to applied H. It is also to be emphasized that all routine H driven process simultaneously exist as they do for any normal AFM under the influence of H. However, canting is spontaneous and therefore once the magnetization state is prepared by an *in-field cooling* (or heating) cycle, as long as history effects are not wiped off by heating the sample above AFM to paramagnetic transition (~ 950 K), an ambiguity in the data (of the order of differences in FC/ZFC) is likely to exit. This also depends on the magnitude of H used in the previous run, due to the presence of *time-stable* $\mu$.

**Weak Ferromagnetism & *Time-stable* $\mu$** : As we have recently shown, manifestation of canting appears in the form of *time-stable* $\mu$ with a functional form, evidently different from other physical mechanisms.[21, 28] From the measurements conducted on single crystal α-$Fe_2O_3$, we found that this *time-stable* $\mu$ appears in WFM region and vanishes in pure AFM region and the effect is also significantly tunable by nano scaling [21]. More importantly, while M increases with increasing H, the corresponding $\mu$ exhibits a peak like behavior with increasing H[21]. In case of remnant state prepared following a FC or ZFC cycle, the net magnetization depends on the Zeeman energy as well as the contribution of average net FM moment corresponding to the spontaneously canted WFM domains in the direction of H. If the remnant state is prepared with H above a critical value, the Zeeman energy dominates. In this case, after switching off H, the remanence decays instantaneously. Between the two extremes, magnetization dynamics is governed by net FM contribution from` spontaneously canted spins and the Zeeman Energy. This leads to a peak like pattern in H dependence of *time-stable* $\mu$ as we have observed in a number of canted AFM [21]. We have also shown experimentally that once a remnant state is set due to H applied in a certain direction, removing H or reversing its direction in WFM region does not alter the magnitude or the direction of remanence.[21] The robustness of this remanence is indicative of WFM domains in the direction of H and flipping its direction would mean flipping of super exchange driven AFM sublattice, which is highly energetically unfavourable.[11] This also explains the *time*-stable character of $\mu$.

Thus the presence of this *time-stable* $\mu$, exclusively associated with canting explains the ambiguity in magnitude of M (and PzM) on repeated cooling[13, 38] and the importance of the exact sequence in which FC/ZFC runs are recorded **(Figures S2)**. The above discussion holds for both type of samples presented in this work. Taking into account all these factors, we put an upper limit to the ambiguity related with the magnitude of M, which can be ~ 0.1 - 0.2 emu/g in case of template or bare $\alpha$-$Fe_2O_3$ nano particles reported earlier[21].

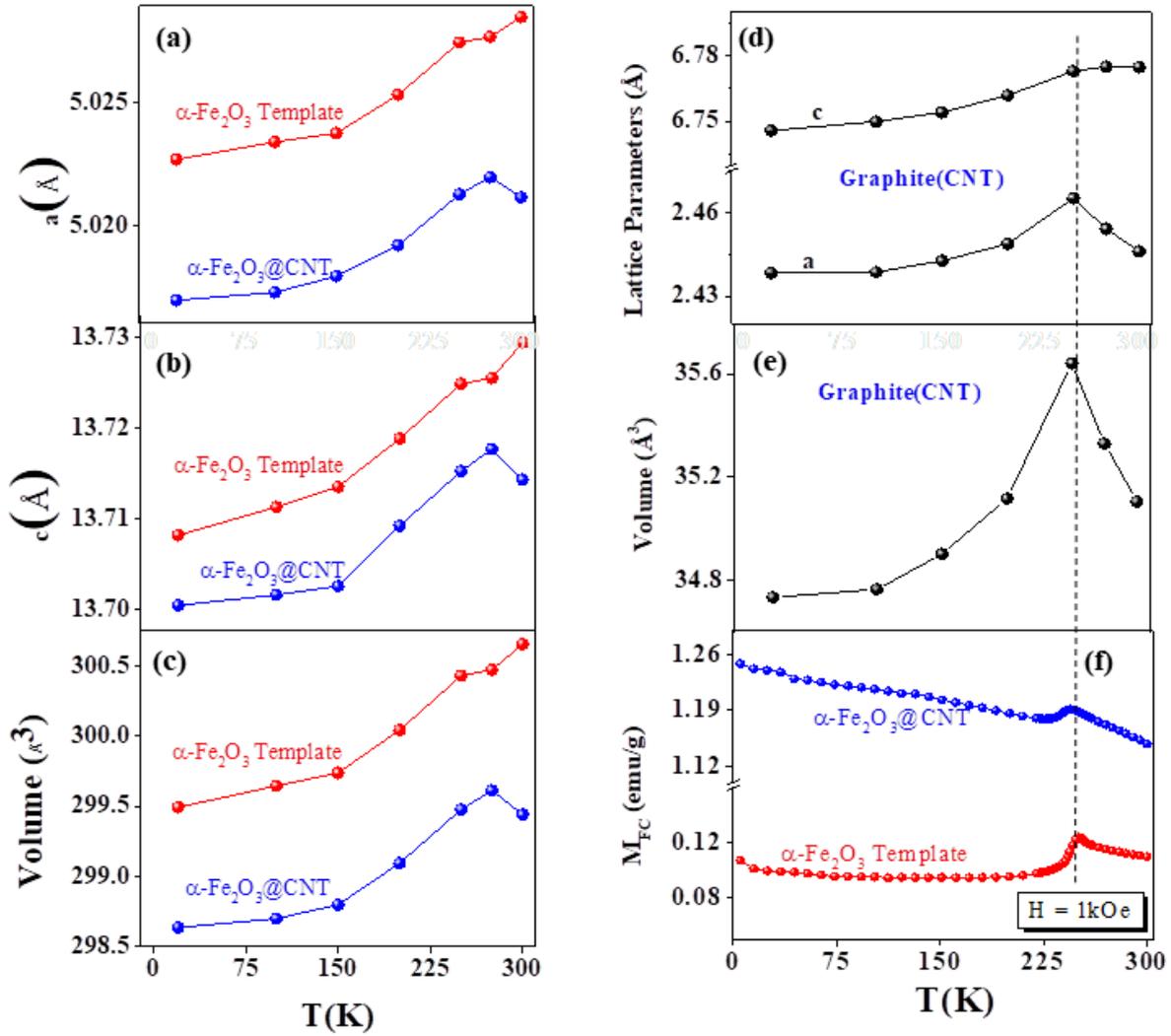

*Figure 3:* *(a)* and *(b)* compare the temperature variation of lattice parameters "a" and "c" for α-Fe$_2$O$_3$@CNT (blue dots) and the template (red dots). *(c)* Volume of the unit cell, exhibiting slight anamoly near ~250K. *(d)* lattice parameters of CNT in the sample α-Fe2O3@CNT. Here, lattice parameter "a" of CNT exhibit a clear anomaly at ~ 250K, around the Morin Transition of the encasulate. This is evident from the M vs T data presented in *(f)* for the bare particles of α-Fe2O3 ( red dots) and α-Fe2O3@CNT (blue dots). The dotted black line is a guide to eye.

From **Figure 2a,** it is also evident that both samples are governed by similar physical process as the data for α-Fe$_2$O$_3$@CNT is a scaled version of the template. However, for the observation of about order of magnitude enhancement in the magnitude of M as well as μ in α-Fe$_2$O$_3$ sample cannot be explained by the ambiguity related to heating / cooling protocols which is typically ~ 0.1 emu/g as estimated from the difference (M$_{FC}$-M$_{ZFC}$) as estimated from FC/ZFC cycles at any fixed H (**Figure 2**). Thus, the enhancement in M as observed in the α-Fe$_2$O$_3$@CNT as compared to the template is much larger than what can be explained by history effects. Therefore, we conclude that the enhancement in magnitude of M and correspondingly μ in α-Fe$_2$O$_3$@CNT is not fully due to either nano scaling (both contain nano wires of α-Fe$_2$O$_3$) or morphology of α-Fe$_2$O$_3$ (which is quite similar in both samples). This prompted us to explore the oxide/CNT interface and associated strain effects arising due to lattice mismatch in more details.

**Temperature variation of Lattice Parameters across T$_M$** : We have recently shown the correlation between structural parameters and the magnitude of for a number of symmetry allowed DMI driven canted AFMs.[21] The c/a ratio also exhibits a much rapid fall with decreasing temperature in case of MnCO$_3$ and exhibits a more pronounced anomaly in the WFM region [21] as compared to α- Fe$_2$O$_3$. Here, MnCO$_3$ is known to be a stronger WFM / PzM as compared to α-Fe$_2$O$_3$ due to its lower Neel transition temperature.[2, 10,14] In view of these observations, we investigated the temperature variation of lattice parameters for both α-Fe$_2$O$_3$@CNT as well as the template. For the lattice parameters, the synchrotron XRD data of α-Fe$_2$O$_3$@CNT sample has been fitted using two phase model in Rietveld Profile Refinement,[32] corresponding to rhombohedral α-Fe$_2$O$_3$ (*R -3 c*; hex-setting) and the CNT (*P 63 m c*).[39] In case of the template, all peaks are identified with α-Fe$_2$O$_3$ (**Figure S1**). For both the samples, lattice parameters corresponding to α-Fe$_2$O$_3$ phase as well as the respective volume of its unit cell are compared in **Figure 3a-3c**. Significant compression effects are observed in the entire temperature range (20-300K) for both the "*a*" and "*c*" lattice parameters of α-Fe$_2$O$_3$@CNT (blue dots) as compared to its bare nano particles (red dots). A slight anomaly in the lattice parameters of α-Fe$_2$O$_3$ around T$_M$ is seen in both the samples. In this aspect, the data are similar to that observed in bare α-Fe$_2$O$_3$ nano crystals, formed using hydrothermal method[21]. Surprisingly we find a pronounced feature in the lattice parameter "*a*" of the graphitic phase (corresponding to CNT), which coincides with the Morin transition $T_M$, intrinsic to the oxide encasulate (**Figure 3d**) . A slight anamoly at this temperature also exists for the lattice parameter "*c*" corresponding to the CNT (**Figure 3d-3f**). This implies that the SO coupling associated with the spin canting phenomenon of the oxide encapsulate is transmitted to the graphitic shells, particularly to its in plane lattice parameter "*a*". This leads to this significant enhancement in the magnetic properties associated with WFM phase near the room temperature. Though microscopic measurements[40,43] are certainly needed to confirm this, it appears that the huge enhancement in the magnitude of magnetization as well as remanence is arising due to this interface effect, which primarily exists between the lattice parameters of α-$Fe_2O_3$ and the CNT. It is also to be recalled that due to curved surface of CNT, as compared to graphene, the SO coupling is considered to be larger.[29] There also have been recent reports on interfacial DMI on graphene/ferromagnetic metal based heterostructures.[44]. However, encapsulation of a symmetry allowed DMI driven canted AFM inside CNT, such as shown here, clearly augments the effect as is evident from **Figures 2-3**.

As mentioned before, larger $T_N$ implies stronger AFM super exchange, leading to smaller spin canting in α-$Fe_2O_3$. Thus, both WFM and PzM effects are relatively smaller in bare α-$Fe_2O_3$ ($T_N$ ~ 950K) as compared to $MnCO_3$ ($T_N$ ~ 30K).[2, 3, 10,14] This feature is reflected in their respective *time-stable* μ as well as temperature variation of their respective lattice parameters.[21] However, the magnitude of μ observed in α-$Fe_2O_3$@CNT at 300K (this work) is now at par with that observed in $MnCO_3$ below 30K.[21] The anomaly in lattice parameters near $T_M$ in case of graphitic shells in α-$Fe_2O_3$@CNT (**Figure 3e**) is also more pronounced than what is observed in the lattice parameters of pure $MnCO_3$ in its WFM phase.[21] Thus, we propose that encapsulation of α-$Fe_2O_3$ within CNT leads to strain and interface effects, which, in turn, modulate the spin canting angle, Fe-O-Fe bond angle and bond lengths[2,3]. The strain effects, as evident from volume compression (**Figure 3c**), are likely to modulate the WFM phase as this can affect both the spin canting angle and a Fe-O-Fe bond lengths/bond angles associated with AFM super exchange.[2, 3] The enhancement in both the magnetization and the remanence is also consistent with these strain effects which are significantly large for α-Fe$_2$O$_3$@CNT.

**Table 1:** *Numerical values of M$_{ZFC}$ and µ$_{ZFC}$ for the sample α-Fe$_2$O$_3$@CNT at the room temperature to depict the magnetization retention.*

| Field (Oe) | T(K) | M$_{ZFC}$ (emu/g) | µ$_{ZFC}$ (emu/g) | Magnetization Retention (%) (µ$_{ZFC}$ / M$_{ZFC}$)*100 |
|---|---|---|---|---|
| 100 | 300 | 0.46413 | 0.32024 | 69 % |
| 1000 | 300 | 1.48752 | 0.59932 | 40 % |
| 10000 | 300 | 2.5721 | 0.2344 | 9 % |

**Weak Ferromagnetism & *Time-stable* µ** : Thus, the highlight of the present work is the *time-stable* remanence, exclusively arising from the weak ferromagnetic phase[21, 45-47], which has been enhanced by an order of magnitude by encapsulation of α-Fe$_2$O$_3$ inside CNT. Looking at µ as the capacity of magnetization retention, two important aspects are its (i) magnitude and (ii) holding time. Since room temperature is more relevant for practical applications, we tabulate the M and µ$_{ZFC}$ data at 300 K for the sample α-Fe$_2$O$_3$@CNT (**Table 1**). Up to 70% of *in-field* M is retained (in the form of µ) after removal of H ~ 100 Oe and 40 % in case of H ~ 1000 Oe for the sample α-Fe$_2$O$_3$@CNT. These H values can easily be achieved by bar magnets. After removal of H, the magnetization decays with time ~ 0.5 % in the measurement time span of two hours α-Fe$_2$O$_3$, whereas it is ~ 5 % in case of the template (**Table 1, Figure 2**). The remanence is stable and larger at 300K than at 5K, further confirming the magnetization dynamics in weak ferromagnetic region is different from other magnetic systems including spin glasses and other nano scale AFM.[17,20] . This effect is observed at room temperature, and therefore it holds promising technological implications. For instance, in routine FM/AFM exchange bias[48] systems, replacing antiferromagnet with a weak ferromagnet should provide a new and a very robust magnetization pinning. This pinning requires rather low magnetic fields and should also be tunable by stress.

In conclusion, appearance of *time stable* remanence is intimately related to Dzyaloshiskii Moriya Interaction driven spin canting phenomenon in *weak ferromagnets* such as *Hematite*. Encapsulation of thsi oxide inside carbon nanotubes provides novel interface effects that lead to much significant enhancement in the magnitude of the *time-stable* remanence at the room temperature. Thus encapsulation inside carbon nanotubes appears to be the most efficient way to manipulate spin canting and associated *weak ferromagnetism*. It is to be emphasized that the method of encapsulation of this multifunctional oxide inside carbon nanotubes is cost effective and scalable and provides ease of integration for direct patterning into spintronic devices.

**Supplementary Material :** Text **S1-S2** ; Figures **S1-S2**

**Acknowledgements**

The authors thank Dr. Sunil Nair (SQUID measurements) and Dr Mukul Kabir (IISER- P) for useful discussions, Dr. A.K.Nigam and Mr. J.Parmar (TIFR) for TEM measurements; Mr. A. Shetty for SEM measurements.AB thanks support from Department of Science and Technology (DST) for Ramanujan fellowship.The authors thank DST and Saha Institute of Nuclear Physics, India for facilitating the experiments at the Indian Beamline, Photon Factory, KEK, Japan.

**References**

[1]   A.P. Balan, S. Radhakrishnan, C.F.Woellner, S.K.Sinha, L.Deng, C.Reyes, B.M.Rao, M.Paulose, R.Neupane, A.Apte et al.. Nature Nanotechnology, **13**, 602 (2018).

[2]   I. E. Dzyaloshinskii, JETP**,** 32, 1259, (1957**)**

[3]   T. Moriya, *Phys. Rev.* **120**, 91 (1960).

[4]   U.K. Roβler, A.N. Bogdanov, and C.P Pfleiderer, Spontaneous skyrmion ground states in magnetic metals, *Nature*, ***442***, 797 (2006).

[5]   V. Baltz, A. Manchon, M. Tsoi, T. Moriyama, T. Ono and Y. Tserkovnyak, *Rev. Mod. Phys.* **90**, 015005, (2018).

[6]   G. Beutier, S.P.Collins, O.V.Dimitrova, V.E. Dmitrienko, M. I. Katsnelson, Y. O. Kvashnin, A. I. Lichtenstein, V. V. Mazurenko, A. G. A. Nisbet, E. N. Ovchinnikova, and D. Pincini, *Phys. Rev. Lett.*, ***119*,** 167201, (2017).

[7]   O. Gomonay, V. Baltz, A. Brataas and Y.Tserkovnyak, *Nature Physics*, ***14*,** 213 (2018).

[8]   L. M. Sandratskii, *Phys. Rev. B* , ***96***, 024450, (2017).

[9]   T. Kikuchi, T. Koretsune, R. Arita & G. Tatara, *Phys. Rev. Lett.*,, 116, 247201 (2016).

[10]  I. E. Dzyaloshinskii,  The problem of piezomagnetism , *JETP**,** **33***, 807 (1957).

[11]  V.P. Andratskiĭ and A.S. Borovik-Romanov, *JETP***, 24,** 687 (1967)

[12]  A.S. Borovik-Romanov, Ferroelectrics *Ferroelectrics***, 162**, 153 (1994). .

[13]  A.S. Borovik-Romanov, *Sov. Phys. JETP* , **11**, 786 (1960).

[14]  V.P. Andratskiĭ, and A.S. Borovik-Romanov, *JETP* , **687**, 1036 (1966)

[15]  S. Blundell,  Magnetism in Condensed Matter,  (Oxford Univ. Press, Oxford, **(2001**).

[16]  F.Donati et.al, , *Science*, ***352****(6283)*, 318 (2016).

[17]  K. Binder, and A.P. Young, *Rev. Mod. Phys.*, **58**, 801 (1986).

[18]  M.J. Benitez, O. Petracic, H. Tüysüz, F. Schüth, and H. Zabel, *Phys. Rev. B* , **83,** 134424, (2011)

[19] M. Suzuki, I.S. Suzuki, and M. Matsuura, *Phys. Rev. B* **2006,** *73*, 184414, (2006)

[20] L. Lundgren, P. Svedlindh, P. Nordblad, and O. Beckman, *Phys. Rev. Lett.*, *51*, 911, (1983)

[21] N. Pattnayak, A. Bhattacharyya, A.K. Nigam, S.W. Cheong and A. Bajpai. *Phys. Rev. B*, **96,** 104422, (2017)

[22] J. Chakhalian, J.W. Freeland, A.J. Millis, C. Panagopoulos, J.M. Rondinelli, *Rev. Mod. Phys.* **86,** 1189, (2014)

[23] D. Halley, N. Najjari, H. Majjad, L. Joly, P. Ohresser, F. Scheurer, C. Ulhaq-Bouillet, S. Berciaud, B. Doudin, and Y. Henry, *Nature Communications*, **5**, 3167 (2014).

[24] R. Saito, G. Dresselhaus, and M.S. Dresselhaus, *Physical Properties of Carbon Nanotubes*. London: Imperial College Press **1998**.

[25] P.M. Ajayan, O. Stephan, Ph. Redlich, and C. Colliex, *Nature*, **375**, 564 (1995).

[26] J.P. Cleuziou, W. Wernsdorfer, T. Ondarcuhu, and M. Monthioux, *ACS Nano* 5, 2348 (2011).

[27] A. Bajpai, S. Gorantla, M. Löffler, S. Hampel, M.H. Rümmeli, J.Thomas, M. Ritschel, T. Gemming, B. Büchner, and R. Klingeler, *Carbon*, **50,** 1706, (2011)

[28] A. Bajpai, Z. Aslam, S. Hampel, R. Klingeler, and N. Grobert, *Carbon*, **114,** 291 (2017)

[29] G.A. Steele, F. Pei, E.A. Laird, J.M. Jol, H.B. Meerwaldt, and L.P.Kouwenhov, *Nature Communications*, **4**, 1573 (2013)

[30] Y. Efroni, S. Ilani, and E. Berg, *Phys. Rev. Lett.*, **119**, 147704, (2017)

[31] A. Kapoor, N. Singh, A.B. Dey, A.K. Nigam, and A. Bajpai, *Carbon*, **132**, 733 ( 2018).

[32] R.A. Young, (International Union of Crystallography. Oxford University Press, (1993).

[33] Ö. Oezdemir, D.J. Dunlop and T.S. Berquo, Geochemistry, Geophysics Geosystems, an electrical journal of earth sciences, *Geochem., Geophys. Geosys.* **9**, Q10ZO1.(2008)

[34] S. Mitra, S.Das, S.Basu, P.Sahu, and K.Mandal, Journal of Magnetism and Magnetic Materials , *321*, 2925 (2009).

[35] P.A. Joy, P.S. Anil Kumar, and S.K. Date, *Journal of Physics: Condensed Matter*, *10,* 11049.(1998)

[36] B.C. Zhao, Y.Q. Ma, W.H. Song, and Y.P. Sun. *Phys. Lett. A* **2006**, *354*, 472 92006).

[37] D.J. Dunlop, *Journal of Geophysical Research*, *112***,** B11103, (2007).

[38] J. Sandonis, J. Baruchel, B. Tanner, G. Fillion, V. Kvardakov and K. Podurets, *J. Magn. Magn. Mater.*, *104***,** 350 (1992)

[39] T. Peci, and M. Baxendale, *Carbon*, *98*, 519, (2016).

[40] A. Michels, D. Mettus, D. Honecker, and K.L. Metlov, *Phys. Rev. B*, *94*, 054424, (2016).

[41] J. Zemen, Z. Gercsi, and K.G. Sandeman, *Phys. Rev. B* , *96*, 024451 (2017).

[42] V. E. Dmitrienko, E. N. Ovchinnikova, S. P. Collins, G. Nisbet, , G. Beutier, Y. O. Kvashnin, V. V. Mazurenko, A. I. Lichtenstein, and M. I. Katsnelson, *Nature Physics* *10*, 202 (2014)

[43] A. Kleibert, A. Balan, R. Yanes, P.M. Derlet, C. A. F. Vaz, M. Timm, A. Fraile Rodríguez, A. Béchè, J. Verbeeck, R. S. Dhaka, et al., *Phys. Rev. B*, *95*, 195404, (2017).

[44] H. Yang, G. Chen, A. A. C. Cotta, A. T. N'Diaye, S. A. Nikolaev, E. A. Soares, W. A. A. Macedo, K. Liu, A. K. Schmid, A. Fert and M. Chshiev. *Nature Materials* **,** *17*, 605, 2018.

[45] A.Bajpai, R Klingeler, N. Wizent, A K Nigam, S-W Cheong and B Buechner,  J. Phys. Cond. Matt., *22*, 096005, (2010).

[46] A. Bajpai A and A.K Nigam, *US Patent* 7276226.

[47] A. Bajpai1, P. Borisov, S. Gorantla, R. Klingeler, J. Thomas, T. Gemming, W. Kleemann and B. Buechner, EPL, *91*, 17006, (2010).

[48] W. H Mieklejohn and C. P Bean, *Phys. Rev.* B, *105,* 904, (1957).

# Supplementary Material

## Enhanced Magnetism & Time - Stable Remanence at the Interface of Hematite and Carbon Nanotubes

Aakanksha Kapoor[1], Arka Bikash Dey[2], Charu Garg[1] and Ashna Bajpai[1-3]

1. Department of Physics, Indian Institute of Science Education and Research, Pune-411008, India.

2. Saha Institute of Nuclear Physics, 1/AF Bidhannagar, Kolkata, India.

3. Centre for Energy Science, Indian Institute of Science Education and Research, Dr. Homi Bhabha Road, Pune, Maharashtra-411008, India.

Email: ashna@iiserpune.ac.in

## Text S1

***Synthesis of α-$Fe_2O_3$@CNT and α-$Fe_2O_3$ Template***

The precursor, *Ferrocene* (98% pure), has been procured from Sigma-Aldrich. The synthesis of the samples being studied involves two steps. (i) Synthesis of Iron-filled multi-walled carbon nanotubes (Fe@CNT) and, (ii) Oxidation of the as-prepared Fe@CNT to form α-$Fe_2O_3$@CNT (or α-$Fe_2O_3$ Template) by suitable annealing treatments as described below.

***Fe@CNT****:* The iron-filled CNT (Fe@CNT) are synthesized using solid-state Chemical Vapor Deposition (CVD) method which comprises of pyrolysis of powder ferrocene[31]. The samples being studied here are prepared by sublimation of ferrocene at 300°C, and further pyrolysis at 900°C, where deposition of the CNT takes place. Variations in synthesis parameters is seen to effect the extent of filling within the core cavity and number of residual particles outside the CNT as well as the morphology of sample. The present synthesis parameters have been optimized for the sample formation in aligned forest morphology such as shown in **Figure 1**. The details of the synthesis procedure of Fe@CNT using a single zone furnace are reported in reference 31.

*α-Fe$_2$O$_3$@CNT*: The as-prepared Fe@CNT is converted into α-Fe$_2$O$_3$@CNT by annealing under carbon dioxide atmosphere at 500°C for 20 minutes. Both time and temperature of annealing crucially effect the quality of the graphitic shell and its interface with the oxide encapsulate. The magnitude of enhanced M as well as μ depends on the annealing temperature and time, which in terun controls the quality of the oxide-graphitic interface.

*α-Fe$_2$O$_3$ Template*: α-Fe$_2$O$_3$ Template is formed by further annealing α-Fe$_2$O$_3$@CNT under carbon dioxide atmosphere at a higher temperature, 700°C, for 20 minutes, to intentionally burn the CNT, forming the bare oxide template.

**Text S2**

It is to be noted that a common practice is to heat the sample above T$_N$, prior to another FC/ZFC cycle in a different H. This is not practically possible in the case of α-Fe$_2$O$_3$ with T$_N$ ~ 950 K. A common practice is to first record ZFC cycle during warming, followed by cooling the sample in same H, leading to FC cycle. However, this protocol leads to spin canting guided from H from below T$_M$ during a ZFC run. Hence, in this work, the sample is heated in zero H from below its T$_M$ prior to any M vs T in presence of H (either FC or ZFC). This protocol leads to more reproducible data on repeated temperature cycling in same H. Remnant states prepared in different H are also prepared following this protocol.

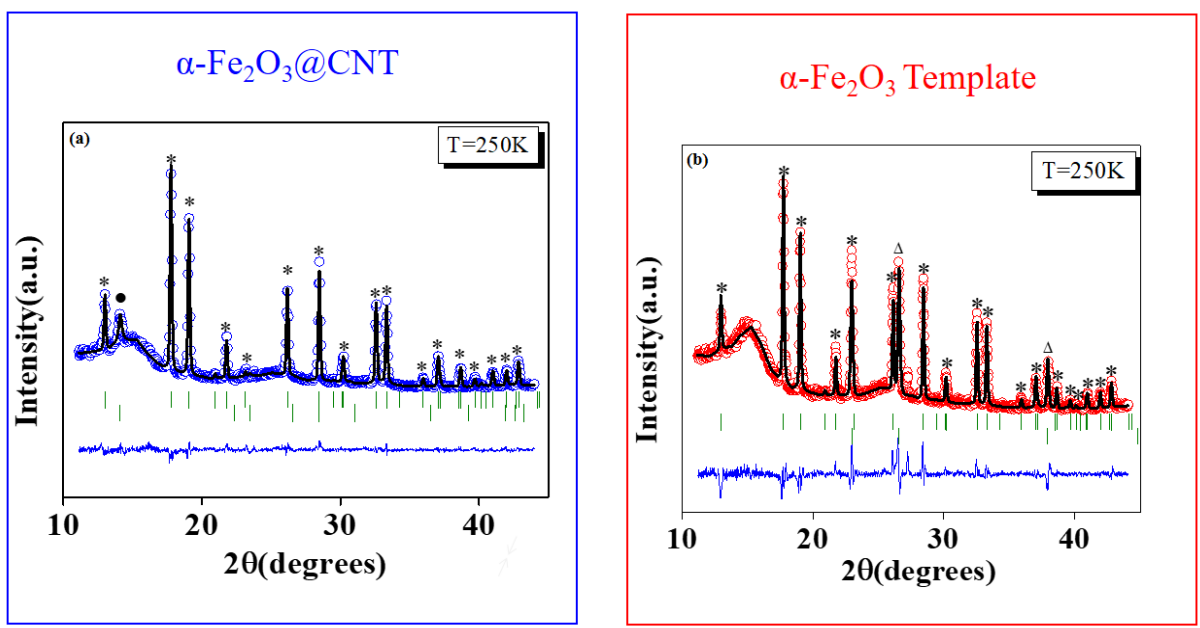

***Figure S1:*** *(**a**) Representative Synchrotron XRD data for α-Fe$_2$O$_3$@CNT. The stars depict Bragg peaks corresponding to α-Fe$_2$O$_3$ phase. The solid black dot indicates Bragg peak corresponding to the graphitic phase (CNT) for the sample α-Fe$_2$O$_3$@CNT . The XRD data has been fitted using two phase model in Rietveld Profile Refinement in this case. (**b**) Synchrotron XRD data for α-Fe$_2$O$_3$ template along with Rietveld fitting (Black line). The stars represent Bragg peaks corresponding to α-Fe$_2$O$_3$. The triangles are from the sample holder. No Bragg peak corresponding to the graphitic phase (CNT) has been observed in this sample.*

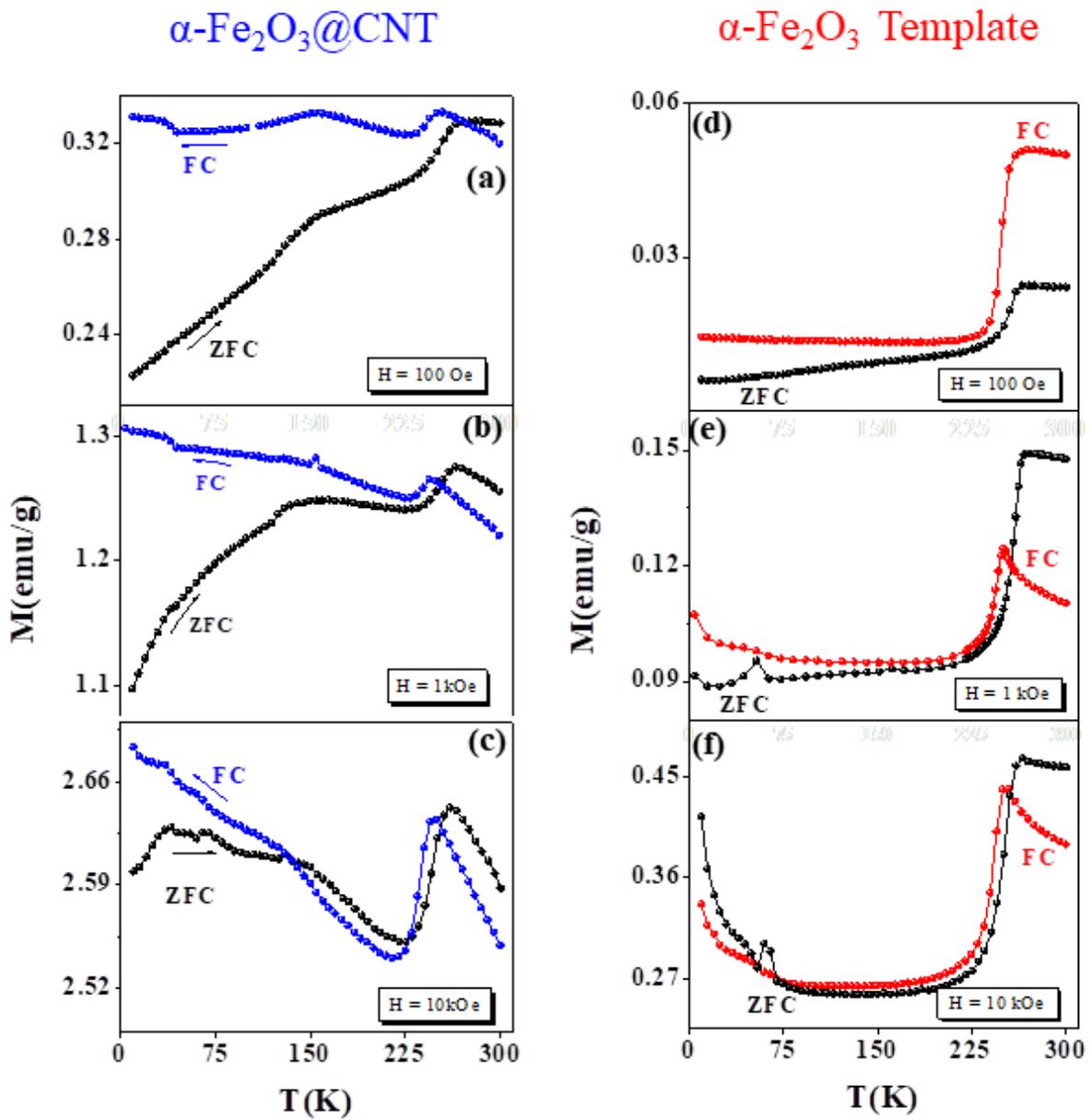

**Figure S2:** *(a) - (c) show M vs T in a typical FC (blue dots) and ZFC (black dots) for the sample α-Fe$_2$O$_3$@CNT measured at 100 Oe, 1kOe and 10 kOe. (d) - (f) show the same for the oxide template. At each measuring H, the magnitude of M is substantially is larger for the sample α-Fe$_2$O$_3$@CNT as compared to what is observed for the oxide template.*